\documentclass{ws-procs975x65}

\def\beq{\begin{equation}}
\def\eeq{\end{equation}}

\begin{document}

\title{Testing the Kerr Paradigm with X-ray Observations}

\author{Cosimo Bambi}

\address{Center for Field Theory and Particle Physics and Department of Physics, Fudan University,\\
Shanghai 200433, China\\
Theoretical Astrophysics, Eberhard-Karls Universit\"at T\"ubingen,\\
T\"ubingen 72076, Germany\\
E-mail: bambi@fudan.edu.cn}

\begin{abstract}
Astrophysical black hole candidates are thought to be the Kerr black holes of general relativity, but the actual nature of these objects has still to be confirmed. The continuum-fitting and the iron line methods are currently the only available techniques to probe the spacetime geometry around these bodies and test the Kerr black hole paradigm. The continuum-fitting method is a robust approach, but the shape of the disk's thermal spectrum is in general too simple to measure the spin and to constrain possible deviations from the Kerr solution at the same time. The iron line analysis is potentially a powerful technique, but at the moment we do not have high quality data and a robust astrophysical model.
\end{abstract}


\bodymatter


\section{Introduction}

Black hole candidates are astrophysical compact objects that can be naturally interpreted as the Kerr black holes of general relativity and that could be something else only in the presence of new physics\cite{narayan}\footnote{For instance, the maximum mass for a neutron star is not more than 3~$M_\odot$ for any reasonable matter equation of state. If a compact object in a binary has a mass exceeding this limit, it is automatically classified as a black hole candidate.}. However, general relativity has been tested only in weak gravitational fields, mainly with experiments in the Solar System and observations of binary pulsars, while the Kerr black hole paradigm relies on the validity of the theory in the strong gravity regime. The observational confirmation of the Kerr nature of black hole candidates would thus represent an important test of Einstein's theory of gravity.

Tests of the Schwarzschild solution in the weak field limit are conveniently performed within the PPN (Parametrized Post-Newtonian) approach\cite{will}. In this framework, one writes the most general static and spherically symmetric metric as an expansion in $M/r$ 
\beq\label{eq-ppn}
ds^2 = - \left( 1 - \frac{2M}{r} + \beta \frac{2M^2}{r^2} + . . . \right) dt^2
+ \left( 1 + \gamma \frac{2M}{r} + . . . \right) \left( dx^2 + dy^2 + dz^2 \right) \, ,
\eeq
where the term $2M/r$ in $g_{tt}$ is to recover the correct Newtonian limit, while $\beta$ and $\gamma$ are two coefficients that parametrize our ignorance. In general relativity, the only spherically symmetric vacuum solution is the Schwarzschild metric and, when cast in the form~(\ref{eq-ppn}), $\beta = \gamma = 1$. To test the Schwarzschild metric, we employ the line element in~(\ref{eq-ppn}) and we determine $\beta$ and $\gamma$ from observations. Current constraints are\cite{will}
\beq
|\beta - 1| < 2.3 \cdot 10^{-4} \, , \quad
|\gamma - 1| < 2.3 \cdot 10^{-5} \, , 
\eeq
and confirm the Schwarzschild solution in the weak field limit within the precision of today measurements.

With a similar approach, one can try to test the Kerr metric. The starting point is a general line element in which a number of ``deformation parameters'' are used to quantify possible deviations from the Kerr geometry. The value of these deformation parameters must be determined from observations. The Kerr black hole hypothesis is confirmed if astrophysical measurements require the Kerr metric in which all the deformation parameters vanish. For the time being, there is not a general formalism as the PPN one, and the problem is that we want to test the gravitational field close to BH candidates, so it is not possible to perform an expansion in $M/r$. There are some proposals in the literature and a common choice is the Johannsen-Psaltis metric\cite{jp}.

The X-ray spectrum of black hole candidates has features that depend on the spacetime geometry around these objects and therefore the analysis of these features can test the Kerr metric\cite{review}. Today there are only two main approaches to probe the spacetime geometry around black hole candidates: the study of the thermal spectrum of thin accretion disks (continuum-fitting method)\cite{cfm0a,cfm0b} and the analysis of the iron K$\alpha$ line\cite{iron0a,iron0b}. Both the techniques were originally developed to measure the black hole spin under the assumption of the Kerr background, but they can be extended to test the Kerr metric\cite{cfm1,cfm2,iron}.

\section{Continuum-fitting method}

The continuum-fitting method can only be applied to stellar-mass black hole candidates. The temperature of the disk scales as $M^{-1/4}$. In the case of a stellar-mass compact object accreting at 10\% the Eddington rate, the spectrum is in the soft X-ray band and good observations are possible. For supermassive black hole candidate of millions or billions Solar masses, the spectrum falls in the UV/optical band where dust absorption makes an accurate measurement impossible.

In the Kerr metric, the spectrum is determined by five parameters: the black hole mass $M$, the black hole spin parameter $a_*$, the mass accretion rate $\dot{M}$, the distance of the source $d$, and the inclination angle of the disk $i$. If we can obtain independent estimates of $M$, $d$, and $i$, we can fit the continuum and infer the black hole spin parameter $a_*$ and the mass accretion rate $\dot{M}$. Without independent measurements of $M$, $d$, and $i$, this is not possible because there is a fundamental degeneracy among the parameters of the model.

\begin{figure}
\begin{center}
\hspace{-0.6cm}
\includegraphics[type=pdf,ext=.pdf,read=.pdf,width=7.0cm]{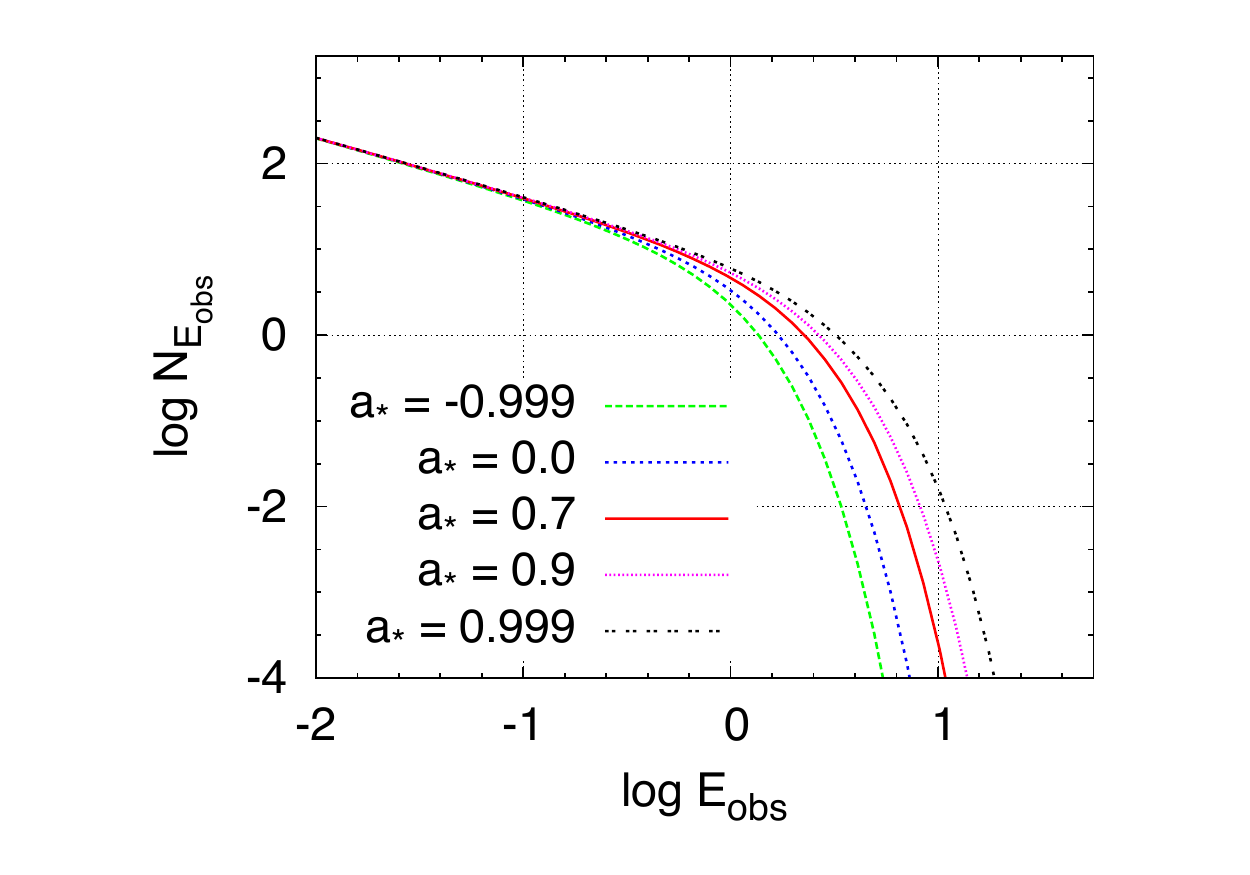}
\hspace{-1.2cm}
\includegraphics[type=pdf,ext=.pdf,read=.pdf,width=7.0cm]{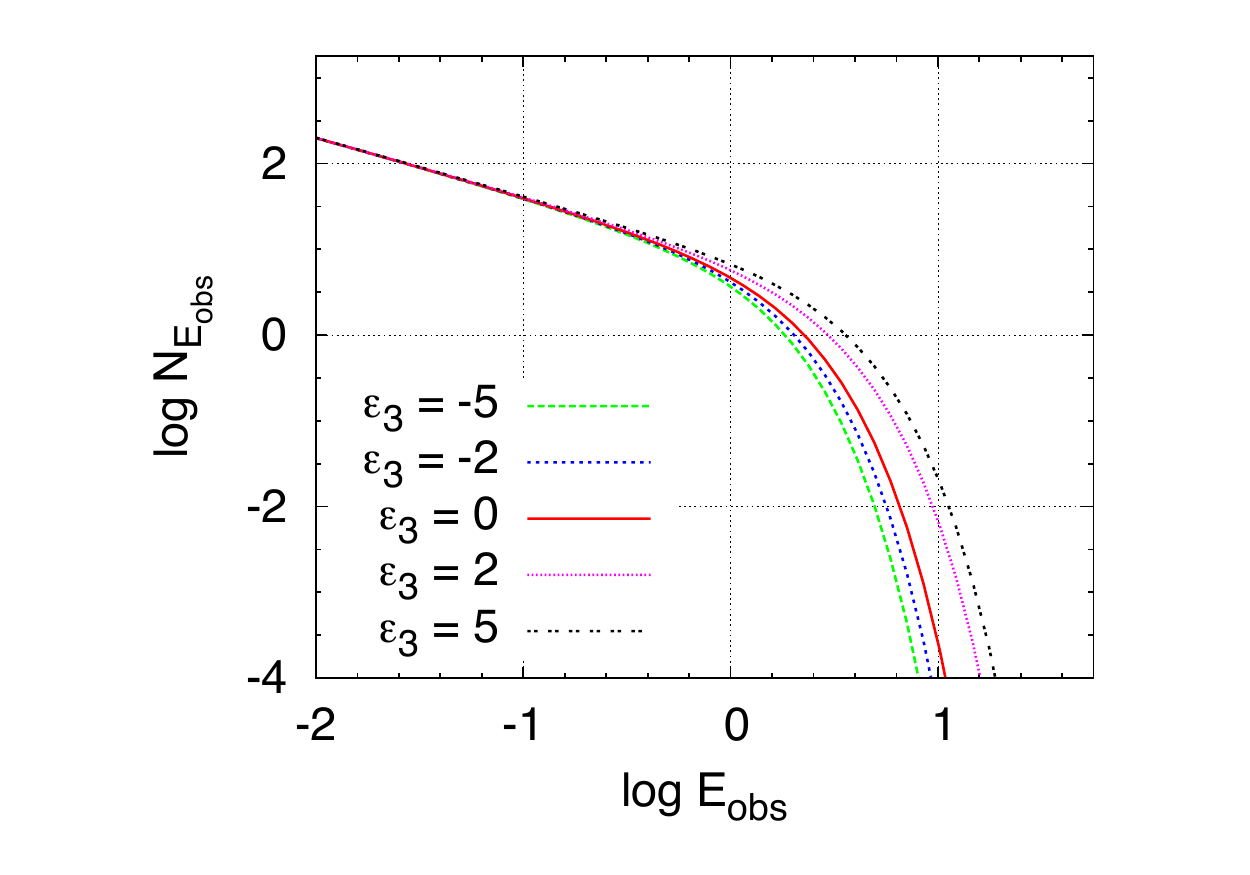}
\end{center}
\vspace{-0.4cm}
\caption{Impact of the spin parameter $a_*$ (left panel) and of the Johannsen-Psaltis deformation parameter $\epsilon_3$ (right panel) on the thermal spectrum of a thin disk. \label{fig1}}
\vspace{0.8cm}
\begin{center}
\hspace{-1.2cm}
\includegraphics[type=pdf,ext=.pdf,read=.pdf,width=7.0cm]{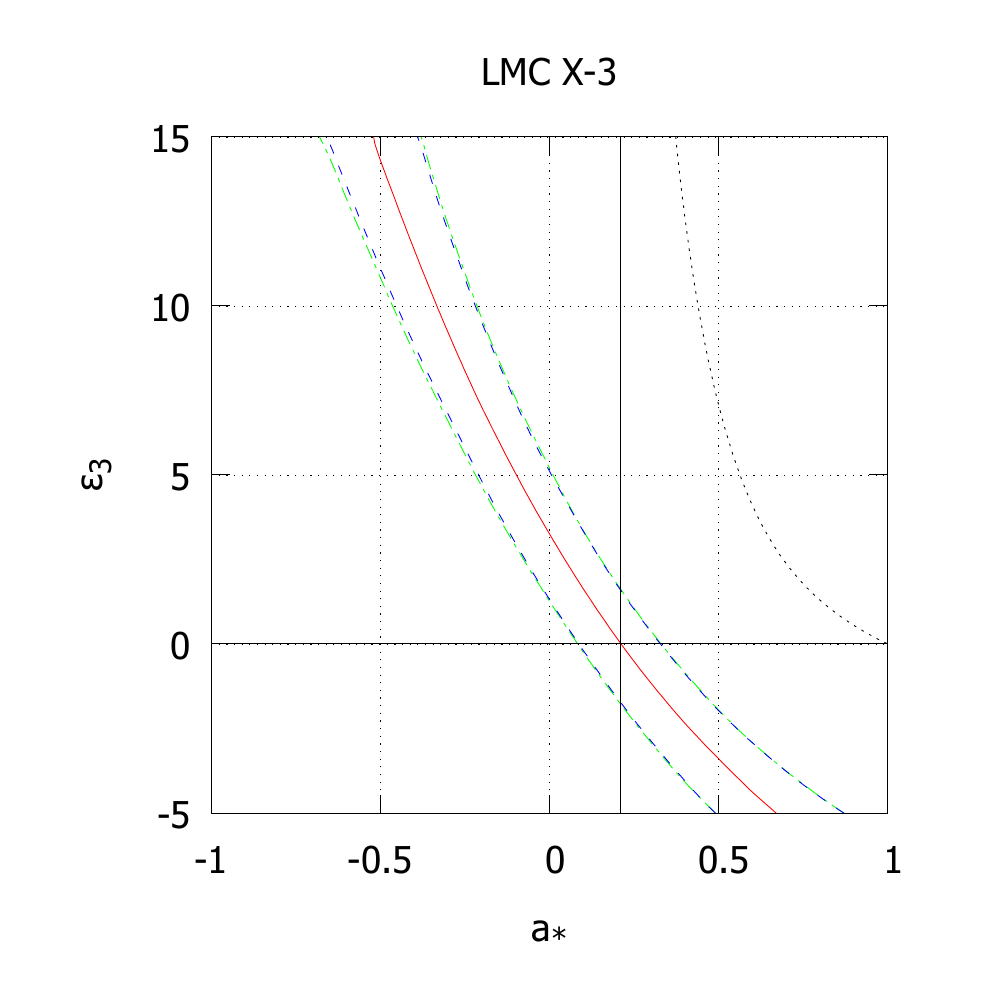}
\hspace{-0.5cm}
\includegraphics[type=pdf,ext=.pdf,read=.pdf,width=7.0cm]{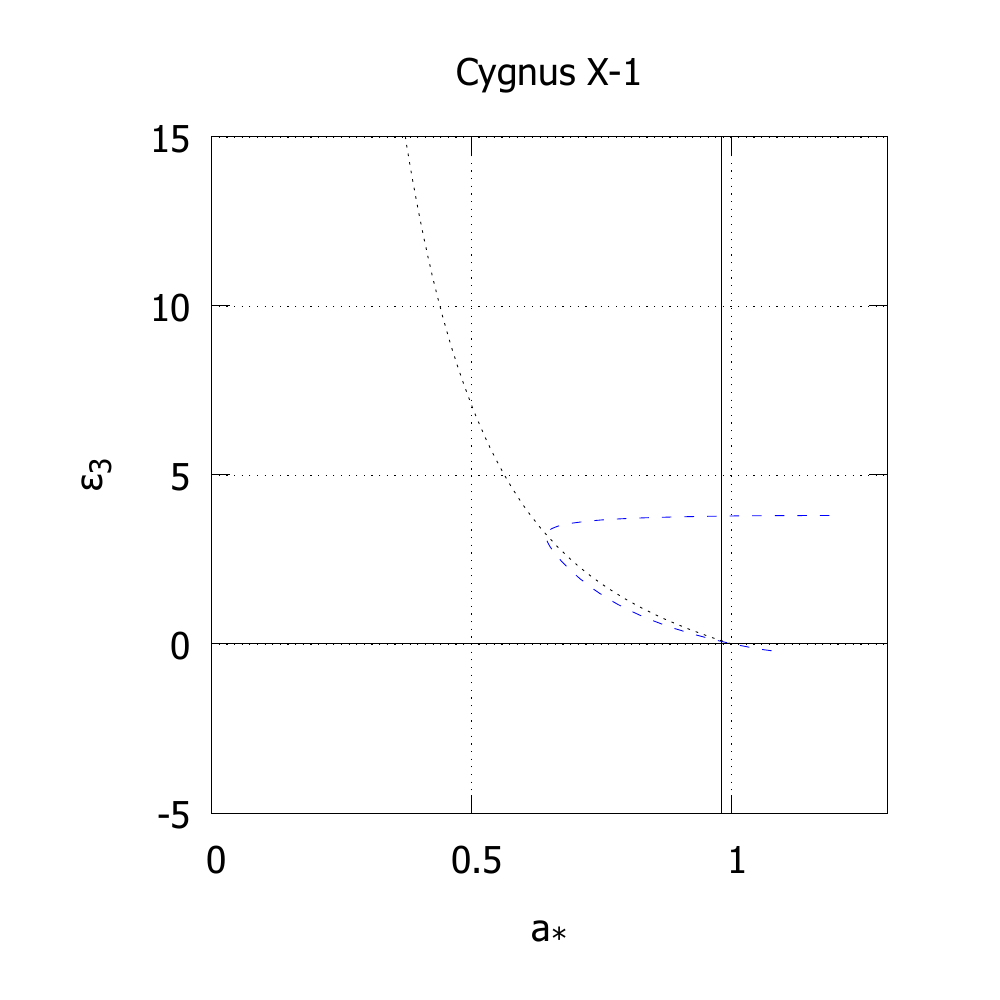}
\hspace{-1.2cm}
\put(-55,87){\small Allowed Region}
\put(-100,115){\small Excluded Region}
\end{center}
\vspace{-0.4cm}
\caption{Continuum-fitting constraints on the Johannsen-Psaltis parameter $\epsilon_3$ from LMC~X-3 (left panel) and Cygnus~X-1 (right panel). From Ref.~13. See the text for more details. \label{fig2}}
\end{figure}

In we want to use the continuum-fitting method to test the Kerr metric, we need to introduce at least one deformation parameter. The impact of the spin parameter $a_*$ and of the Johannsen-Psaltis deformation parameter $\epsilon_3$ on the thermal spectrum of a thin disk is shown in Fig.~\ref{fig1}. Even without a quantitative analysis, it is clear that the effect of $a_*$ and $\epsilon_3$ is very similar, namely they both move the high energy cut-off of the spectrum to lower or higher energies. The point is that the disk radiates as a blackbody locally, and as a multi-color blackbody when integrated radially. The position of the high energy cut-off is essentially determined by the disk's radiative efficiency, which is just one number. If we assume the Kerr metric, there is a one-to-one correspondence between the value of the radiative efficiency and the inner edge of the disk, which is at the innermost stable circular orbit and only depends on $a_*$. If we relax the Kerr black hole hypothesis and we consider the possibility of a non-vanishing deformation parameter, then the radiative efficiency depends on both the spin and possible deviations from the Kerr solution.

The result is that in general we cannot constrain the spin and the deformation parameter at the same time. We typically find an allowed region on the spin parameter - deformation parameter plane as that in the left panel in Fig.~\ref{fig2}. The spacetimes along the solid red line generate the same spectrum. However, if a black hole candidate looks like a very fast-rotating Kerr black hole, it may be possible to get a constraint on the deformation parameter as shown in the right panel in Fig.~\ref{fig2}. The reason is that compact objects much more oblate or prolate than a Kerr black hole have an innermost stable circular orbit at larger radii and therefore a disk with a lower radiative efficiency. On the contrary, a thin disk around a fast-rotating Kerr black hole has a high radiative efficiency. This is true only for some non-Kerr metrics\cite{cfm-b,lingyao}, while other types of deformations cannot be constrained\cite{cfm-cpr}. The continuum-fitting method constraints on the Johannsen-Psaltis parameter $\epsilon_3$ for 10 stellar-mass black hole candidates are reported in Ref.~\refcite{lingyao}.

\section{Iron line spectroscopy}

The iron K$\alpha$ line observed in the X-ray spectrum of black hole candidates is thought to be generated by the illumination of the accretion disk by a hot corona above the black hole. While there are also other emission lines in the spectrum, the iron K$\alpha$ line is particularly strong. Moreover, this line is very narrow in frequency in the rest-frame of the emitter, while that observed in the spectrum of black hole candidates is broad and distorted, and its shape can be explained as the result of relativistic effects. The analysis of the profile of the iron K$\alpha$ line can thus provide information on the spacetime metric around the compact object. The technique can be used for both stellar-mass and supermassive black hole candidates.

\begin{figure}
\begin{center}
\hspace{-0.6cm}
\includegraphics[type=pdf,ext=.pdf,read=.pdf,width=7.0cm]{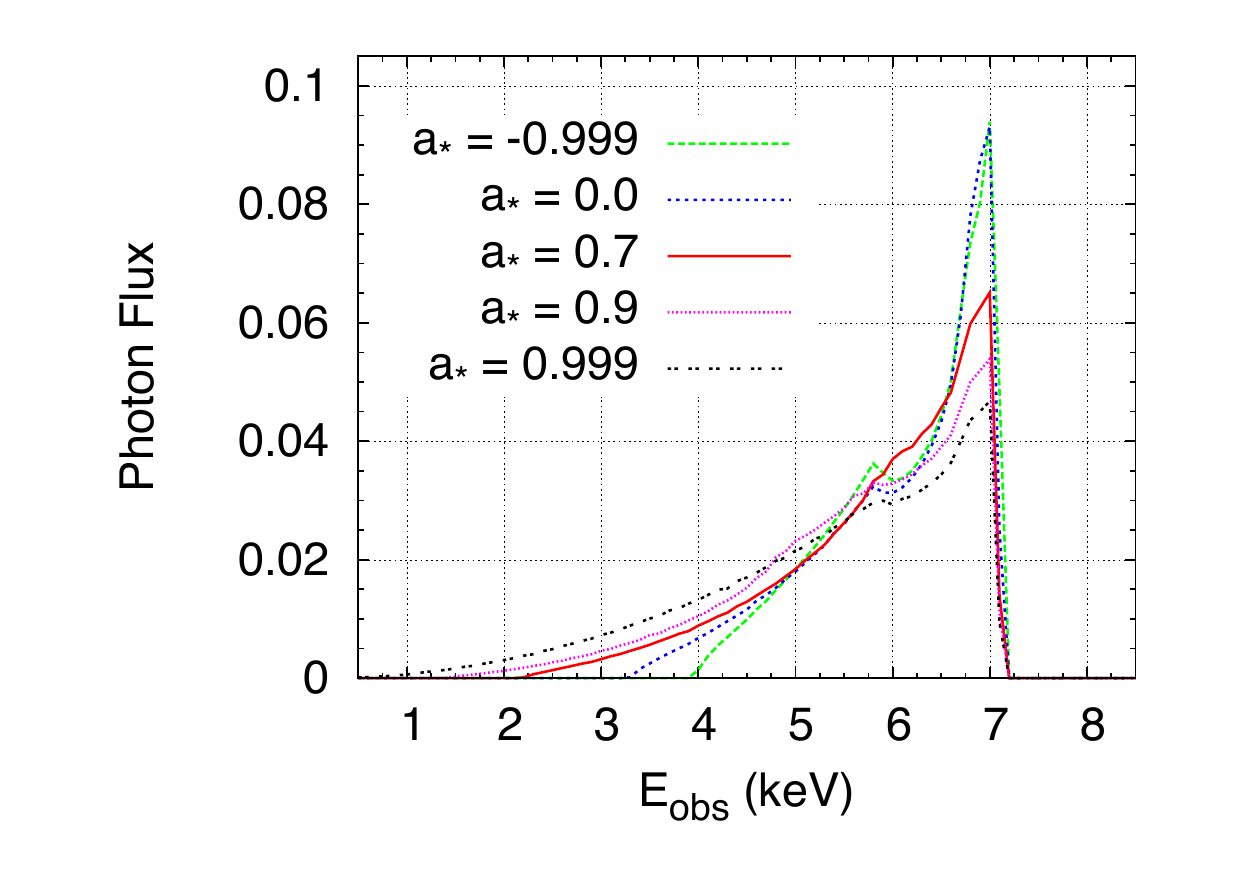}
\hspace{-1.2cm}
\includegraphics[type=pdf,ext=.pdf,read=.pdf,width=7.0cm]{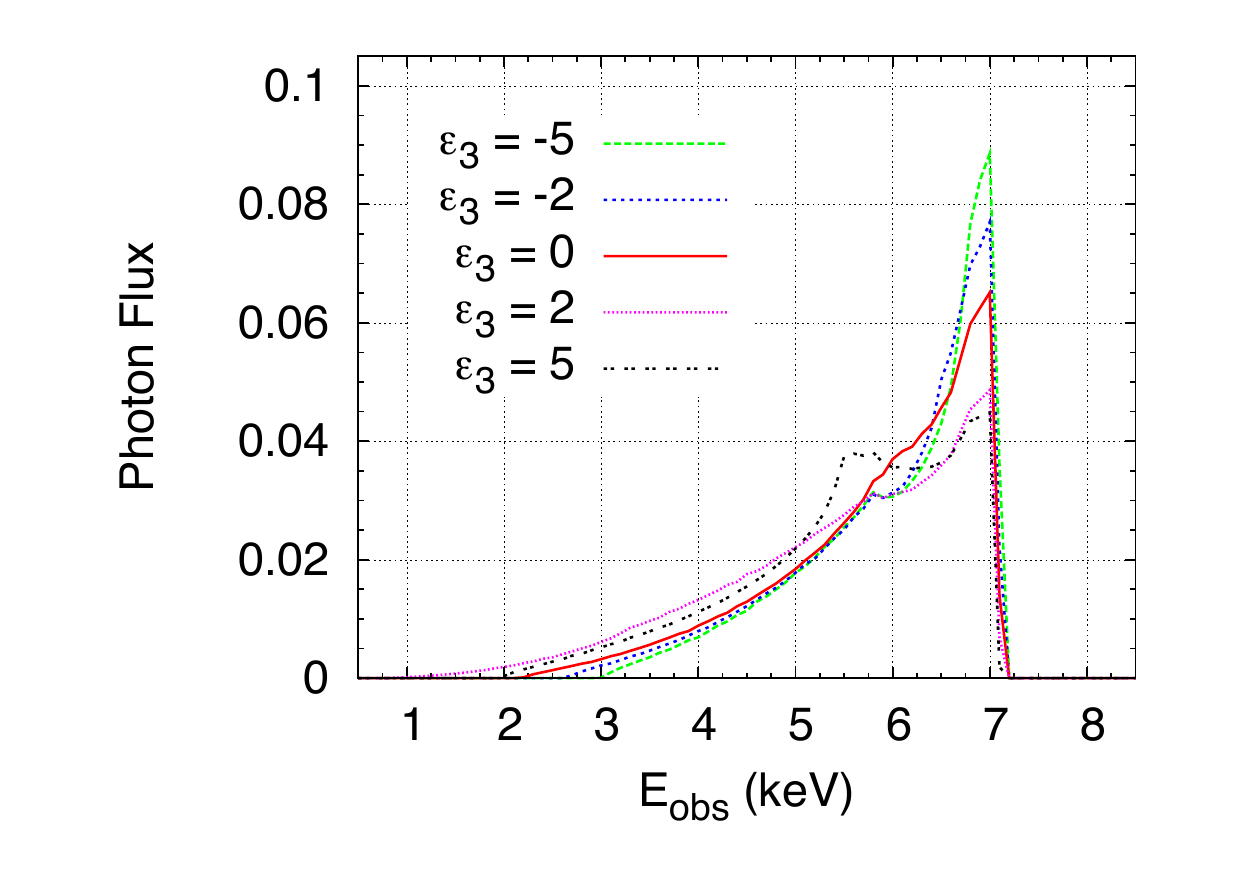}
\end{center}
\vspace{-0.4cm}
\caption{Impact of the spin parameter $a_*$ (left panel) and of the Johannsen-Psaltis deformation parameter $\epsilon_3$ (right panel) on the iron line profile. \label{fig3}}
\end{figure}

The shape of the iron line has a quite complicated structure resulting from the radiation emitted from different radii. The impact of the the spin $a_*$ and of the Johannsen-Psaltis deformation parameter $\epsilon_3$ on the iron line profile is shown in Fig.~\ref{fig3}. As found in Refs.~\refcite{jjc1} and \refcite{jjc2}, the analysis of the iron line can potentially break the degeneracy between the estimate of the spin and possible deviations from the Kerr metric, even if some deformations parameters are definitively more challenging to constrain than others\cite{jjc2}. However, there are two issues. First, high quality data are necessary. With current X-ray facilities, a good observation of a bright AGN may have up to $10^3$ photons in the iron line, but in general this is not enough to break the degeneracy between the spin and the deformation parameter\cite{jjc1}. With current data, the constraining powers of the continuum-fitting and the iron line methods are probably comparable. There is some work in progress on the analysis of X-ray data to make this point more clear (Jiang et al., in preparation). Second, any measurement is as good as the theoretical model employed in the measurement. At present, there are several uncertainties: some fundamental assumptions of the model (e.g. that the inner edge of the disk is at the innermost stable circular orbit) are still to be verified, the geometry of the corona is not know, and the emissivity profile (which is a crucial ingredient in the iron line measurement) is currently modeled with a simple power law.

Current observations of the iron line in the spectrum of black hole candidates can anyway test and rule out some exotic scenarios. For instance, boson stars or other equilibrium configurations maintained by matter with an exotic equation of state can typically be ruled out as black hole candidates\cite{exotic1}. The gravitational force around similar objects is never strong enough to significantly redshift the radiation emitted at small radii. The result is that the iron line profile in these spacetimes cannot have the characteristic low energy tail, which is expected in the case of fast-rotating Kerr black holes and is observed in the spectrum of a few stellar-mass black hole candidates and of several supermassive black hole candidates. Other exotic black hole alternatives, like traversable wormholes, are more difficult to test and rule out as a general class, but they may be tested in specific cases\cite{exotic2}.

\section{Concluding remarks}

The continuum-fitting method seems to be able to provide quite reliable measurements, but the shape of the thermal spectrum of a thin disk is simple and in general it is fundamentally impossible to measure the spin and constrain possible deviations from the Kerr metric at the same time. The observation of objects that look like fast-rotating Kerr black holes can anyway constrain some types of deformations.

The analysis of the iron line profile is potentially a powerful tool to test the Kerr metric. However, one needs a high photon count number to reduce the intrinsic noise of the source, as well as a good theoretical model to fit the data, in particular a good understanding of the emissivity profile. Today the photon number in the iron line is limited, the exact geometry of the corona is not known, and there are only phenomenological models for the emissivity profile.

It is worth noting that there are a few stellar-mass black hole candidates for which (assuming the Kerr metric) the spin parameter has been measured by both the continuum-fitting and the iron line methods. These measurements typically agree. However, this cannot confirm the Kerr nature of black hole candidates, because this is what one would expect even if these objects were not Kerr black holes, but just from the assumption that the inner edge of the disk is at the innermost stable circular orbit\cite{cfm-iron}. However, the agreement between the two methods can provide support on the validity of the iron line measurements.


\section*{Acknowledgments}

This work was supported by the NSFC grant No.~11305038, the Shanghai Municipal Education Commission grant No.~14ZZ001, the Thousand Young Talents Program, Fudan University, and the Alexander von Humboldt Foundation.



\begin{thebibliography}{0}

\bibitem{narayan}
R.~Narayan, {\em New J.\ Phys.}\  {\bf 7}, 199 (2005).

\bibitem{will} 
C.M.~Will, {\em Living Rev.\ Rel.}\  {\bf 9}, 3 (2006).

\bibitem{jp}
T.~Johannsen and D.~Psaltis, {\em Phys.\ Rev.\ D} {\bf 83}, 124015 (2011).

\bibitem{review} 
C.~Bambi, {\em Mod.\ Phys.\ Lett.\ A} {\bf 26}, 2453 (2011).

\bibitem{cfm0a} 
S.N.~Zhang, W.~Cui and W.~Chen, {\em Astrophys.\ J.}\  {\bf 482}, L155 (1997).

\bibitem{cfm0b} 
J.E.~McClintock, R.~Narayan and J.F.~Steiner, {\em Space Sci.\ Rev.}\  {\bf 183}, 295 (2014).

\bibitem{iron0a} 
A.C.~Fabian, M.J.~Rees, L.~Stella and N.E.~White, {\em MNRAS}\  {\bf 238}, 729 (1989).

\bibitem{iron0b} 
C.S.~Reynolds, {\em Space Sci.\ Rev.}\  {\bf 183}, 277 (2014).

\bibitem{cfm1} 
C.~Bambi and E.~Barausse, {\em Astrophys.\ J.}\  {\bf 731}, 121 (2011).

\bibitem{cfm2} 
C.~Bambi, {\em Astrophys.\ J.}\  {\bf 761}, 174 (2012).

\bibitem{iron} 
C.~Bambi, {\em Phys.\ Rev.\ D} {\bf 87}, 023007 (2013).

\bibitem{cfm-b} 
C.~Bambi, {\em Phys.\ Lett.\ B} {\bf 730}, 59 (2014).

\bibitem{lingyao} 
L.~Kong, Z.~Li and C.~Bambi, {\em Astrophys.\ J.}\  {\bf 797}, 78 (2014).

\bibitem{cfm-cpr} 
C.~Bambi, {\em Phys.\ Rev.\ D} {\bf 90}, 047503 (2014).

\bibitem{jjc1} 
J.~Jiang, C.~Bambi and J.F.~Steiner, {\em JCAP} {\bf 1505}, 025 (2015).

\bibitem{jjc2} 
J.~Jiang, C.~Bambi and J.F.~Steiner, arXiv:1504.01970 [gr-qc].

\bibitem{exotic1} 
C.~Bambi and D.~Malafarina, {\em Phys.\ Rev.\ D} {\bf 88}, 064022 (2013).
  
\bibitem{exotic2} 
C.~Bambi, {\em Phys.\ Rev.\ D} {\bf 87}, 084039 (2013).  

\bibitem{cfm-iron}
C.~Bambi, {\em JCAP} {\bf 1308}, 055 (2013).

\end{thebibliography}
\end{document}